\documentclass [12pt,aps]{article}
\begin{document}

\title{BF THEORY ON A BRANE}

\author{ D. L. Henty\\
{\it Department of Physics and Astronomy},\\ {\it
University Of British Columbia,} \\ {\it Vancouver, British Columbia
V6T 1Z1 Canada }\\email: henty@physics.ubc.ca}
\maketitle

\begin{abstract}

An alternative approach to introducing gravitational dynamics on a
brane embedded in a higher dimensional spacetime is presented. The
brane is treated as a boundary of a higher dimensional manifold in
which the bulk action is described by a metric independent topological
quantum field theory. The example of a five dimensional non-Abelian BF
theory with a boundary brane is considered. A natural boundary condition is
adopted chosen for consistency of the topological action despite the presence
of a boundary. The resulting effective action on the brane is the action of general
relativity in first order form plus terms involving the extrinsic
curvature of the brane.

\end{abstract}

\newpage

\section{Introduction}

The idea of a brane embedded in large extra dimensions as an
alternative to Kaluza-Klein compactification has received considerable
attention since a specific model to achieve this was proposed in
[1]. In this approach matter fields are confined to
 the brane as open
strings ending on the brane while gravitational modes are confined
close to the brane by the unique characteristics of the gravitational
field of a brane within a five dimensional anti-deSitter spacetime. In
[1]  the action of general relativity was taken to govern the
gravitational dynamics of the five dimensional spacetime. This choice
for the gravitational action was motivated as a low energy effective
theory of an underlying string theory. A metric was therefore
introduced from the outset. On the other hand, the idea has frequently
been expressed that in string theory, or indeed in any fundamental
theory of quantum gravity, the metric should really be a derived
feature [2]. Therefore, one is motivated to consider the possible
consistency of a brane embedded in a spacetime of large extra
dimensions without the initial assumption of a metric.

Consistent metric independent theories are of course known in the form
of topological quantum field theories [3]. Therefore, one approach to
exploring the consistency of a brane embedded in a metric free
spacetime of large extra dimensions would be to consider a brane
embedded in a higher dimensional topological field theory. This has a
collateral benefit in that there would be no gravitational modes
propagating in the bulk since a topological field theory has no local
dynamical modes. In such an approach several problems immediately
present themselves, however. First of all, it is not clear how a brane
having local dynamics could couple to a topological quantum field
theory without spoiling its consistency. Secondly, there must be a
natural way for general relativity to emerge on the brane from the
topological theory in the bulk. That is, general relativity cannot
simply be introduced on the brane in an ad hoc manner.

In approaching these problems, the consistency issue is most easily
examined if the brane is considered as a boundary of a one higher
dimension manifold on which is defined a specific topological field
theory. For example, a 3 brane may be taken to enclose a finite four
volume or the 3 brane may divide an infinite four volume into two half
spaces with the other boundary at infinity. The question of
consistency then becomes whether a topological theory and a boundary
condition can be found in which the consistency of the topological
theory in the bulk is maintained and which allows dynamics on the
boundary. Topological field theories on manifolds with boundary have
been considered by a number of authors [4]-[10]. The easiest way to
ensure consistency of the bulk topological theory when introducing a
boundary is to simply impose a boundary condition where all fields
vanish at the boundary.  However, this prevents any dynamics from
emerging on the boundary. Several authors have considered the
possibility of other consistent boundary conditions which can also
introduce local dynamics on the boundary. In the most extensively
studied such approach dynamical ``edge states'' are introduced on the
boundary in a manner chosen to cancel surface terms arising from
variation of the bulk action to thereby maintain a consistent
topological field theory in the bulk [6]-[9]. While very interesting
for a variety of reasons, such edge states do not appear to be able to
provide the dynamics of general relativity. In another approach,
general relativity was introduced on the boundary of a five
dimensional space and then a corresponding topological field theory
was identified in the bulk [10].

In the present paper non-Abelian BF theory on a manifold with a bounndary is considered
with a natural boundary condition in the form of a
normal boundary condition on B, chosen in [5] solely for mathematical
consistency of the bulk topological field theory. The normal boundary condition is
expressed in an equivalent dual formulation which takes advantage of
the metric needed to define the normal and tangential forms at the
boundary. As a specific model, non-Abelian BF theory in a 5D
arbitrarily large metric free manifold with a boundary brane is
considered. The B and F[A] fields are chosen to take values in
SO(4,1). Remarkably, the normal boundary condition on
B is satisfied by a relation between the B field and the soldering
form induced on the boundary which leads to the first order Palatini
action for general relativity on the boundary after integration over
the extra dimension.

Therefore, the obstacles to a consistent embedding of a brane world in
a metric independent topological field theory can be overcome in the
model considered in this paper.

The paper concludes with some speculations on the possibility of branes 
naturally arising in topological field theories as defects in the context of 
topological field theory as an ``intrinsically broken'' gauge field theory.

\section{BF Theory On A Manifold With Boundary}

First the approach of [5] will be briefly reviewed before extending the 
approach to non-Abelian BF theory. The topological field theory considered 
in [5] had the following action:

\begin{equation}
\label{eq1}
S = \int {B\wedge dC} 
\end{equation}

\noindent
where the integral is over an n dimensional manifold M with boundary,
B is a p form and C is an n-p-1 form. To kill the boundary term which
appears in the variation of the action (\ref{eq1}) the following
natural boundary condition was chosen:

\begin{equation}
\label{eq2}
B \in {\mathop {\Omega} \nolimits_{nor}^{p}} 
\end{equation}

\noindent
where ${ \mathop {\Omega} \nolimits_{nor}^{p} } $ is the space of all p-forms 
normal to the boundary of M, i.e., the space of all p-forms which vanish 
when contracted with any tangent vector of the boundary of M. A number of 
features of the above action were then considered in [5] which will not be 
needed here. However, an observation noted in [5] which will be useful here 
is that a metric, which is needed to distinguish tangential and normal forms 
to the boundary, can be used to define the Hodge dual map over M. The Hodge 
dual in turn maps the space of p forms tangential to the boundary to n-p 
forms normal to the boundary (and vice versa). Specifically:

\begin{equation}
\label{eq2.5}
\ast :{ \mathop {\Omega} \nolimits_{\tan} ^{p} } (M)\to{ \mathop {\Omega 
}\nolimits_{nor}^{n - p} } (M)
\end{equation}

\noindent
where ${ \mathop {\Omega} \nolimits_{\tan} ^{p} }(M)$ is the space of p-forms 
tangential at the boundary and * is the Hodge dual defined over M.

The action for Abelian BF theory can be written as follows in arbitrary 
dimension:

\begin{equation}
\label{eq3}
S = \int {B\wedge dA} 
\end{equation}

\noindent
where A is the connection of an Abelian group over M and B is an n-2
form.  Therefore, Abelian BF theory is just a special case of the
topological field theories considered in [5]. To extend the approach
of [5] to non-Abelian BF theory we consider the general BF theory
action:

\begin{equation}
\label{eq4}
S = \int {B\wedge F} 
\end{equation}

\noindent
where F[A] is the curvature two form of the connection A of a
principle bundle over M associated with a non-Abelian group G and B is
a group algebra valued n-2 form where n is the dimension of the
manifold M [3]. Variation of the action results in the following
equations of motion:

\begin{equation}
F = 0  ~~{\rm  and }~~ DB = 0
\label{eq4.5}
\end{equation}

\noindent
where

\begin{equation}
\label{eq5}
DB = dB + [A,B]
\end{equation}

Variation of the action also introduces a surface term on the boundary 
$\partial $M and a boundary condition must be introduced to cancel this 
term. Following the above approach we choose the normal boundary condition 
(\ref{eq2}) to kill the surface term. Because of the above dual map (3) we may 
equivalently write the boundary condition (\ref{eq2}) as:

\begin{equation}
\label{eq6}
B \in \ast { \mathop {\Omega} \nolimits_{\tan} ^{n - (n - 2)} } 
\end{equation}

\noindent
i.e., B is restricted to the space of forms dual to the space of 
tangential two forms.

\section{BF Theory On A World Brane}

We next apply the above approach to consistent boundary conditions in 
non-Abelian BF theory to a 3 dimensional brane evolving inside an 
arbitrarily large 5D manifold M without metric and with the bulk 
action described by non-Abelian BF theory. The four 
dimensional world brane may be considered as a boundary of the 4+1 
dimensional manifold M since there is no metric in the bulk and we can 
without loss of generality identify the two timelike directions in the 
boundary and bulk. Therefore, we can equivalently consider M to be a five 
dimensional manifold with four dimensional boundary brane $\partial $M. We 
take G to be SO(4,1) (the analysis will be essentially the same for 
SO(3,2)). The 3 brane may be taken to enclose a finite four volume in M or 
the 3 brane may divide an infinite four volume into two spaces with the 
other boundary taken at infinity (e.g., with trivial boundary condition with 
all fields vanishing at infinity). The total action is:

\begin{equation}
\label{eq7}
S = k\int {B\wedge F} 
\end{equation}

\noindent
where k is a coupling constant, with the boundary condition at the brane  given by (\ref{eq6}) to kill
the surface terms on the brane, i.e.:

\begin{equation}
\label{eq8}
B \in \ast {\mathop {\Omega} \nolimits_{\tan} ^{2}} 
\end{equation}

Now a natural basis for the space of tangential forms is given by the local 
frame fields corresponding to the induced metric on the brane. Since B also 
takes values in G the tangent basis forms must also take values in 
G=SO(4,1). Therefore, a tangential 2 form basis is given by the wedge product $e^{A}\wedge e^{B}$ (where A,B are group indices). Therefore, a solution 
of (\ref{eq8}) at the boundary brane can be written:

\begin{equation}
\label{eq9}
B = \ast ^{(5)}(e\wedge e)
\end{equation}

\noindent
where *$^{(5)}$ refers to the 5D Hodge dual.

The total 5D action (\ref{eq7}) can be split into two pieces S$_{1}$
and S$_{2}$ where S$_{1}$ is the action for a 5D volume comprising a
thin region at the boundary where the boundary condition holds (e.g.,
brane thickness) and S$_{2}$ is the action in the bulk. Thus, using
the above dual formulation of the boundary condition:

\begin{equation}
\label{eq10}
S = S_{1} + S_{2} = k\int \ast ^{(5)}(e\wedge e) \wedge F + k\int 
{B\wedge F} 
\end{equation}

\noindent
where both integrals are 5D volume integrals. The first term is the effective action at the brane since the second term has no dynamical modes. The first term has a similar form to 5D general relativity in first order form but is not 5D covariant since the frame fields e are purely tangential. The 5D curvature F at the brane can be split into
normal-normal (5-5), normal-tangent (5-i) and tangent-tangent (i-i)
(where i=1-4) components using the metric introduced at the brane to
define the normal boundary condition. The 5-5 and 5-i components of F
get truncated by the contraction with the purely tangential frame
fields $e\wedge e$ leaving only the tangent-tangent components of the
5D curvature F in the first term of (\ref{eq10}). In general the
internal 5 components will not be killed, however. This can be better examined in a component form of the first term of (\ref{eq10}):

\begin{equation}
\label{eq18}
S_{1}=k\int d^{5}x\vert e \vert e^{[i}_{A}e^{j]}_{B}F^{AB}_{ij}=k\int d^{5}x\vert e\vert (e^{[i}_{a}e^{j]}_{b}F^{ab}_{ij}+2e^{[i}_{[5}e^{j]}_{b]}F^{5b}_{ij})
\end{equation}

\noindent 
where $\vert e\vert$ is the magnitude of the determinant of e, a,b=1-4 are group indices in an SO(3,1) subgroup of SO(4,1) and F=$F^{(5)}$ is the 5D curvature.
  The equations of
Gauss and Codazzi may now be used to express the 5D curvature in
terms of the intrinsic curvature F$^{(4)}$ on the brane and the
extrinsic curvature K [13]:

\begin{equation}
\label{eq11}
F^{(5)}_{ij} = F^{(4)}_{ij} + K_{i} K_{j}-K_{j}K_{i}+K_{i;j}-K_{j;i} 
\end{equation}

\noindent
where internal indices have been suppressed.
Inserting this in S$_{1}$ in equation (\ref{eq18}) gives the action at the brane in purely 4D terms:

\begin{equation}
\label{eq12}
S_{1}=k\int d^{5}x\vert e\vert (e^{[i}_{a}e^{j]}_{b}(F^{(4)ab}_{ij}+K^{a}_{i}K^{b}_{j}-K^{a}_{j}K^{b}_{i})+2e^{[i}_{[5}e^{j]}_{b]}(K^{b}_{i;j}-K^{b}_{j;i}))
\end{equation}

\noindent
where $F^{(4)}$ is the 4D curvature but where the integration is still over the 5D volume at the brane 
where the boundary condition holds.

The action (\ref{eq12}) may be reduced to a 4D action in a trivial manner if we assume all fields are independent of the fifth normal 
direction over the local region where the boundary condition holds so that we 
can integrate S$_{1}$ over x$^{5}$. The result is simply (\ref{eq12}) integrated over the 4D volume times a constant of integration t which is just the normalized thickness of the local boundary region where the 
boundary condition holds (for example, t may be the brane thickness, where the brane is viewed as a defect as discussed below). This assumed independence of the action on x$^{5}$ also affects the extrinsic curvature terms. The extrinsic curvature K may be written in
Gaussian normal coordinates as [13]:

\begin{equation}
\label{eq15}
K_{ij} = -\frac {1}{2} \partial g_{ij}/ \partial x^{5}
\end{equation}

With the assumed independence of all quantities on $x^{5}$ the extrinsic curvature will vanish. Equation (\ref{eq12}) then
becomes, removing the explicit reference to 4D and replacing F with the more 
suggestive R:

\begin{equation}
\label{eq14}
S_{1} = k'\int d^{4}x\vert e'\vert e^{[i}_{a}e^{j]}_{b}R^{ab}_{ij}  
\end{equation}

\noindent
(where $\vert e'\vert $ is the magnitude of the determinant of $e^{i}_{a}$, and where $k'$=tk), which is just the action for 4D general relativity in the first order Palatini 
form.

Although the above assumptions resulting in the action (\ref{eq14}) for 4D general relativity are reasonable, it is interesting to look closer at the action (\ref{eq12}). In particular, it is interesting to note that while the equations of motion require F$^{(5)}$ to vanish, F$^{(4)}$ need not due to the extrinsic  curvature terms. That is, the extrinsic curvature acts like a source for the 4D curvature. Thus, starting from a topological BF theory with only flat connections as solutions a dynamical theory with a source emerges from the boundary condition.

\section{Defect Formation In Topological Field Theories}

The present section provides a very qualitative discussion of how a 
world brane could arise as a defect in the context of a bulk 
topological quantum field theory. As noted in the introduction the present 
paper is motivated, at least in part, by the philosophy that the most basic 
formulation of quantum gravity should be metric free. Topological quantum 
field theories provide many attractive aspects for such a metric free 
underlying theory. Topological quantum field theories have typically not 
been considered as viable models for a realistic physical theory due to the 
lack of any realistic symmetry breaking mechanism or other mechanism for 
creating a lower energy theory which possesses local dynamics. The model in 
this paper illustrates how general relativity can emerge naturally on a 
boundary brane of a bulk topological field theory. However, ideally the 
topological field theory should also be able to explain brane formation. 
Since topological field theories have no local dynamics it is not clear how 
such a boundary brane could form in the first place, especially absent any 
mechanism for symmetry breaking.

Topological field theories do inherently possess aspects of spontaneous 
symmetry breaking, however. The characteristic feature of 
spontaneously broken field theories is the existence of distinct degenerate 
vacuum states. A specific one of these vacuum states must be chosen as the 
vacuum for the theory and the symmetry reflecting the label that 
distinguishes the vacua is broken. The same vacuum state need not be chosen 
at every place, however, and different domains characterized by different 
vacua are possible. In particular, where regions are initially causally 
disconnected distinct vacuum domains are expected. It is well known that 
such domains can give rise to domain walls and other topological defects in 
a variety of cosmological scenarios based on spontaneously broken field 
theories. [11]

Topological quantum field theories are metric independent theories, therefore:

\begin{equation}
\label{eq20}
\delta S / \delta g = T^{\mu \nu}  = 0
\end{equation}

\noindent
and hence have vanishing energy for all observables. [3] Also, distinct 
vacuum solutions are known for various theories which have been studied, 
which solutions are labeled by topological properties of the underlying 
manifold. Therefore, topological quantum field theories share the 
characteristic feature of degenerate vacuum solutions possessed by 
spontaneously broken field theories. Nonetheless, despite the presence of 
distinct vacuum solutions the complete theory may not truly have distinct 
vacuum states if the states are mixed in the quantum theory. Typically, 
topological field theories are studied on compact manifolds where such 
mixing does occur. Where a theory with degenerate solutions is defined on an 
infinite manifold, however, mixing does not occur and distinct degenerate 
vacua will be present. [12]

Therefore, a topological quantum field theory defined on an infinite 
manifold intrinsically does possess the characteristic features of a 
spontaneously broken field theory. As a result topological quantum field 
theories intrinsically possess the possibility of giving rise to domain 
walls and other defects without any further breaking. Once a defect is 
present the domain wall will act as a boundary to the region described by 
the topological field theory and, as shown above, the boundary dynamics can 
give rise to realistic physical theories. Therefore, based on this 
``intrinsic symmetry breaking'' effect present in a large class of 
topological field theories such theories without more can potentially 
present models for both defect creation and dynamics on the brane. Thus in 
effect two stages of symmetry breaking occur without the need to introduce 
scalars to break the high degree of symmetry present in topological 
theories. Typically scalars have been considered necessary for symmetry 
breaking since only scalars can avoid breaking Poincare invariance at the 
same time. Topological field theories seem to provide an exception to this 
rule.

Therefore, although the model considered in the present paper is far from a realistic 
model of a fundamental theory, nonetheless it is suggestive of a possible 
road from a fundamental topological quantum field theory to realistic four 
dimensional dynamics.

Another suggestive feature of a topological ``intrinsically broken'' 
fundamental theory comes from an analogy with superconductivity. A domain 
wall dividing two regions of a topological field theory is analogous to a 
superconducting region of broken U(\ref{eq1}) gauge symmetry sandwiched between 
two unbroken normal regions. In type II superconductors flux tubes will 
penetrate the superconducting region and interact with each other. If a topological field 
theory is a candidate for a fundamental theory, perhaps with string theory 
or M theory as a dynamical phase, the analog of the flux tubes would be strings 
of topological phase coupled between parallel branes defining the edges of the domain wall. 
Therefore, string theory could naturally emerge as an effective field theory 
on a brane/defect as a broken phase of a topological field theory in the 
bulk. 

\section{Discussion}

A model with a bulk topological field theory defined on a manifold 
with a boundary brane has been presented which provides dynamics on the 
boundary brane in a natural manner. Specifically, SO(4,1) BF theory with a 
natural boundary condition, expressed in its dual form using the induced 
metric at the boundary, results in the action for general relativity on the 
brane after integration over the extra dimension. This provides an 
alternative to ``edge states'' to introduce dynamics into a topological 
field theory with boundary. Also, an alternative means of suppressing 
gravitational modes propagating off a brane world to that in 
[1] is provided 
by this model. In principle this approach could be extended in a variety of 
ways, such as other groups, other topological field theory actions and 
higher dimensions which allow consistent introduction of a soldering form at 
the boundary. The soldering form requirement is reminiscent of soliton 
solutions of purely internal gauge theories and the boundary 
condition of this paper could be viewed as a consistency condition for the brane to be a TQFT defect. A qualitative discussion of possible defect formation in topological field theories was provided.

\noindent
Acknowledgements:
\noindent
The author would like to thank G. W. Semenoff for many valuable discussions.

\noindent
{\bf References:}

\noindent
[1] L. Randall and R. Sundrum, Phys. Rev. Lett. 83 (1999) 4690

\noindent
[2] E.g., G. T. Horowitz, Commun. Math. Phys. 125 (1989) 417

\noindent
[3] D. Birmingham,M. Blau, M. Rakowski and G. Thompson, Physics Reports 209 
(1991) 129

\noindent
[4] M. Blau and G. Thompson, Phys. Lett. B225 (1991) 535

\noindent
[5] S. Wu, Commun. Math. Phys. 136 (1991) 157

\noindent
[6] A. P. Balachandran and P. Teotonio-Sobrinho, Int. J. Mod. Phys. A 9 
(1994) 1569

\noindent
[7] A. Momen, Phys.Lett.B394 (1997) 269-274

\noindent
[8] S. Carlip, Phys.Rev.D55 (1997) 878-882

\noindent
[9] V. Hussain and S. Major, Nucl. Phys. B 500 (1997) 381

\noindent
[10] R. Brooks, Class.Quant.Grav.14 (1997)L87-L91.

\noindent
[11] Vilenkin et al, Cosmic Srings and Other Cosmological 
Defects (Cambridge University Press 1994)

\noindent
[12] S. Weinberg, Quantum Theory of Fields , Vol. II (Cambridge University 
Press 1996) at p. 163

\noindent
[13] C. Misner, K. Thorne and J. Wheeler, Gravitation, (W. H. Freeman 1973)

\end{document}